\documentclass{aa}  

\usepackage{graphicx}
\usepackage{txfonts}
\usepackage{xcolor}
\usepackage{mhchem}
\usepackage{hyperref}
\newcommand{\rev}[1]{#1}
\begin{document}

   \title{Piecing together formic acid isomerism in dark clouds}
   \subtitle{Detection of \textit{cis}-formic acid in TMC-1\thanks{Based on observations carried out with the Yebes 40m telescope (projects 19A003, 20A014, 20D023, 21A011, 21D005, and 23A024). The 40m radio telescope at Yebes Observatory is operated by the Spanish Geographic Institute (IGN; Ministerio de Transportes, Movilidad y Agenda Urbana).} and astrochemical modeling}

   \author{G. Molpeceres
          \inst{1} 
          M. Ag\'undez
          \inst{1}
          M. Mallo
          \inst{1}
          C. Cabezas,
          \inst{1}
          M. Sanz-Novo,
          \inst{2}
          V.M. Rivilla,
          \inst{2}
          J. Garc\'ia de la Concepci\'on,
          \inst{3}
          I. Jim\'enez-Serra,
          \inst{2}
          J. Cernicharo.
          \inst{1}
          }
   \authorrunning{Molpeceres et al}
   \institute{Departamento de Astrofísica Molecular, Instituto de Física Fundamental (IFF-CSIC), C/ Serrano 121, 28006 Madrid, Spain\\
              \email{german.molpeceres@iff.csic.es}
              \and
              Centro de Astrobiolog{\'i}a (CAB), INTA-CSIC, Carretera de Ajalvir km 4, Torrej{\'o}n de Ardoz, 28850 Madrid, Spain
              \and
              Departamento de Ingeniería Química y Química Física, Facultad de Ciencias, and ICCAEx, Universidad Extremadura, Badajoz, Spain
             }

   \date{Received \today; accepted \today}

 
  \abstract
   {The presence of molecular isomers in interstellar environments has become a topic of growing interest within the astrochemical community. Contrary to predictions based on thermodynamic equilibrium, recent observations reveal a diverse array of high-energy isomers and conformers. One of the most iconic molecular isomers detected in space, formic acid (\ce{HCOOH}, FA), has been the focus of extensive theoretical research aimed at understanding its speciation into cis and trans conformers in dark clouds and photodissociation regions. In this work, we report the detection of c-FA, the higher-energy conformer, using ultrasensitive observations of TMC-1. This detection adds to previous findings in the Barnard-5 and L483 dark clouds. The derived trans-to-cis isomer ratio in TMC-1, 17.5, closely matches those observed in other sources, suggesting that the same chemical processes are at play across these environments. To investigate this, we conducted detailed astrochemical gas-grain models tailored to formic acid isomerism to explain the observed ratios. Our models successfully reproduce the observed trans/cis ratios and indicate that the presence of cis-formic acid can be attributed to the release of c-FA from grains, followed by isomerization driven by the excess energy released during the desorption process, a process that we name as isomerization upon desorption. The models also show that the isomerization of t-FA to c-FA in the gas phase is negligible at 10 K, meaning the observed ratios are a direct consequence of the formation pathways of both isomers on the surface of dust grains. However, at higher temperatures, quantum tunneling mediated direct isomerization in the gas becomes significant, and the ratios converge toward the thermodynamic equilibrium value.}   


      \maketitle
%

\section{Introduction}

One intriguing aspect of interstellar medium (ISM) chemistry is the observed ratios of isomers detected by radio telescopes.\footnote{We use the term isomerism broadly here, encompassing positional isomerism (same molecular formula but different atomic connectivity), stereoisomerism (molecules differing only in their three-dimensional spatial arrangement), and conformerism (a subset of stereoisomerism where isomers interconvert via rotation around a single bond, as in the case of formic acid). } Unlike on Earth, in the ISM many high-energy isomers—whose energy separation would normally preclude their presence in significant amounts—are found to be abundant \rev{\citep[see, for example,][]{Brunken2009, 2009ApJ...690L..27M, Laas2011, Loomis2013, Loomis2015, Cuadrado2016, 2016Sci...352.1449M, Taquet2017, Agundez2018, Agundez2019, Agundez2023, Rivilla2019, Cernicharo2021a, Cernicharo2022, Cernicharo2024, Marcelino2021, Cabezas2021, Rivilla2023, SanAndres2024, 2024A&A...692A...5E, sanz-novo_conformational_2025}}. These detections challenge the notion that the relative stability of the isomers can be used for abundance prediction, a concept known as the minimum energy principle \citep{Lattelais2009, Chataigner2024}. Over the years, an increasing body of observational evidence has shown that most interstellar isomers do not adhere to this principle \citep[e.g.,][]{Loomis2015, Shingledecker2019, Shingledecker2020, GarciadelaConcepcion2021, Molpeceres2022c,Rivilla2023,sanz-novo_conformational_2025}. The reason for this discrepancy is clear: in the ISM, the extremely low temperatures (10 K) impose strict kinetic control over the formation and destruction of molecules, including isomers. Unfortunately, understanding the kinetics of these processes requires explicit knowledge of a vast number of chemical reactions and their associated rate coefficients on gas and on the surface of dust grains \citep[some non-exhaustive examples include][]{Laas2011,Vazart2015, Shingledecker2019,Shingledecker2020,  Ishibashi2021, Ballotta2021, Molpeceres2022c,GarciadelaConcepcion2023}. Since these reactions vary for each molecule and type of isomer, a detailed study of each case is, in many cases, necessary.

Formic acid (HCOOH, FA) is one of the molecules for which our understanding of molecular isomerism in the ISM is most advanced. This molecule presents two different conformational isomers, commonly labeled cis (c-FA) and trans (t-FA). The more stable of the two is t-FA, with a trans-cis gap of approximately 2050 K \citep{GarciadelaConcepcion2022}. It is woth remarking that, if molecules would follow theromdynamic ratios, the abundance of c-FA would be negligible, precluding detection. Formic acid was one of the first interstellar molecules to be detected \citep{Zuckerman1971}, initially in Sgr B2 in its trans form. Surprisingly, the first detection of the cis isomer occurred much more recently \citep{Cuadrado2016} at the Orion Bar photodissociation region (PDR). In their study, \citet{Cuadrado2016} explained the presence of c-FA through a photoswitching mechanism, in which isomer-selective UV pumping allows for trans-to-cis interconversion, reaching a pseudo-equilibrium with a trans-cis ratio of $\sim$3. The detection of c-FA in dark clouds occurred later, first in Barnard-5 \citep{Taquet2017}, then in L483 \citep{Agundez2019}, with this work presenting the third detection, in TMC-1. Several other searches could only provide upper limits, apart from a tentative detection reported by \citet{SanzNovo2023} in the G+0.693-0.027 molecular cloud. The detection of c-FA in dark clouds poses a significant riddle, as photoswitching cannot be invoked as the driving force for the isomerism due to the attenuated interstellar radiation field. This is also evident from the observed t-FA / c-FA ratio of nearly 20 in dark clouds.

The presence of c-FA in the cold ISM has provided an excellent framework for postulating and evaluating isomerization mechanisms in interstellar environments. Over the years, various physical and chemical processes have been investigated to rationalize c-FA abundances in dark clouds. These include tunneling-mediated isomerization \citep{GarciadelaConcepcion2022}, gas-phase acid-base reactions \citep{GarciadelaConcepcion2023}, and isomer conversion through hydrogenation reactions on grains \citep{Molpeceres2022c}. Furthermore, recent studies have shown that FA, regardless of its isomeric form, is primarily formed on the surface of icy dust grains via the formation of the HOCO radical through the \ce{CO + OH -> HOCO} reaction \citep{Molpeceres2023,Ishibashi2024}, where the isomerism of HOCO also plays a crucial role in determining the final isomeric form of FA \citep{Molpeceres2025}. It is therefore evident that understanding the isomeric excess of a molecule as simple as FA remains a complex challenge, requiring a comprehensive understanding of both its formation and destruction pathways.

In this work, we report the first detection of c-FA in TMC-1 using the \textsc{Quijote} line survey and determine the corresponding trans-cis ratio. Encouraged by the consistent ratios found across the three dark clouds where the molecule has been detected, we undertook the task of fully rationalizing the isomeric chemistry of FA in dark clouds. The similarity in ratios strongly suggests a common underlying chemistry. Following this reasoning, we use astrochemical models to provide a conclusive explanation not only for the specific abundance of c-FA in dark clouds but also for the lack of definitive detections in other regions.

\section{Observations} \label{sec:observations}

\subsection{\rev{Description of the observations}}

\begin{table*}
\small
\caption{Observed line parameters of c/t-FA in \mbox{TMC-1}.}
\label{table:lines}
\centering
\begin{tabular}{lccccccc}
\hline \hline
\multicolumn{1}{l}{Molecule} & \multicolumn{1}{l}{Transition} & \multicolumn{1}{c}{$\nu_{\rm calc}$} & \multicolumn{1}{c}{$E_{\rm up}$} & \multicolumn{1}{c}{$T_A^*$ peak} & \multicolumn{1}{c}{$\Delta v$\,$^a$}      & \multicolumn{1}{c}{$V_{\rm LSR}$}      & \multicolumn{1}{c}{$\int T_A^* dv$} \\
& & \multicolumn{1}{c}{(MHz)} & \multicolumn{1}{c}{(K)}       & \multicolumn{1}{c}{(mK)}                   & \multicolumn{1}{c}{(km s$^{-1}$)}  & \multicolumn{1}{c}{(km s$^{-1}$)}  & \multicolumn{1}{c}{(mK km s$^{-1}$)} \\
\hline
t-HCOOH & 2$_{1,2}$-1$_{1,1}$ & 43303.710 & 6.3 & $3.94 \pm 0.12$ & $0.88 \pm 0.03$ & $5.87 \pm 0.01$ & $3.70 \pm 0.01$ \\
t-HCOOH & 2$_{0,2}$-1$_{0,1}$ & 44911.740 & 3.2 & $8.04 \pm 0.14$ & $0.92 \pm 0.03$ & $5.87 \pm 0.01$ & $7.83 \pm 0.10$ \\
t-HCOOH & 2$_{1,1}$-1$_{1,0}$ & 46581.226 & 6.5 & $4.12 \pm 0.19$ & $0.84 \pm 0.03$ & $5.88 \pm 0.01$ & $3.70 \pm 0.12$ \\
c-HCOOH & 2$_{1,2}$-1$_{1,1}$ & 42541.400 & 6.7 & -- & -- & -- & --\,$^b$ \\
c-HCOOH & 2$_{0,2}$-1$_{0,1}$ & 43926.482 & 3.2 & $1.78 \pm 0.16$ & $0.87 \pm 0.08$ & $6.01 \pm 0.04$ & $1.64 \pm 0.14$ \\
c-HCOOH & 2$_{1,1}$-1$_{1,0}$ & 45351.360 & 6.9 & -- & -- & -- & --\,$^b$ \\
\hline
\end{tabular}
\tablefoot{The line parameters $T_A^*$ peak, $\Delta v$, $V_{\rm LSR}$, and $\int T_A^* dv$ and the associated errors are derived from a Gaussian fit to each line profile. $^a$\,$\Delta v$ is the full width at half maximum. $^b$ Line is only marginally detected. No fit is attempted.
}
\end{table*}

\begin{figure}
   \centering
   \includegraphics[width=\columnwidth]{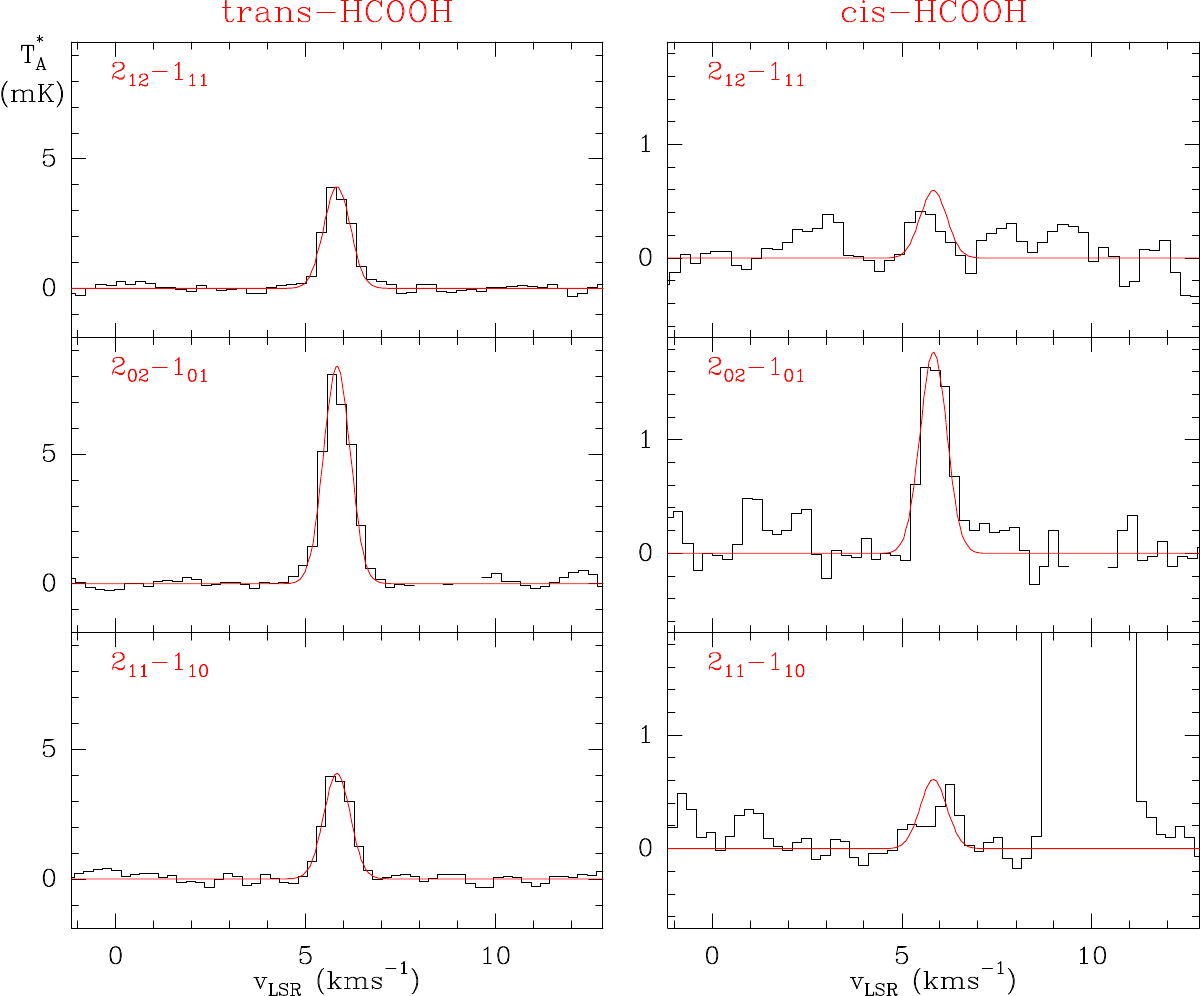}
   \caption{Lines of t-FA and c-FA (left and right panels, respectively) observed in \mbox{TMC-1}. Blanked channels correspond to negative artifacts resulting from the frequency-switching technique. The red lines correspond to the computed line profiles adopting the parameters given in Sect.\,\ref{sec:observations}. \rev{The lines in top and bottom panels of c-FA are only marginally detected (See text)}   }
   \label{fig:line}
   \end{figure}

We used the latest data set of \textsc{Quijote} (Q-band Ultrasensitive Inspection Journey to the Obscure \mbox{TMC-1} Environment; \citealt{Cernicharo2021b}) to search for c-FA in \mbox{TMC-1}. Briefly, \textsc{Quijote} is an on-going line survey carried out with the Yebes\,40m telescope toward the cyanopolyyne peak position of the cold dense cloud \mbox{TMC-1} \rev{($\alpha_{\rm J2000} = 04^{\mathrm{h}}\,41^{\mathrm{m}}\,41.9^{\mathrm{s}},\;
  \delta_{\rm J2000} = +25^{\circ}\,41^{\prime}\,27.0^{\prime\prime}$)}. The observations are carried out using the frequency-switching technique and the data cover the Q band in the frequency range 31.0-50.3 GHz with a spectral resolution of 38.15 kHz. The latest data, already presented in \cite{Cabezas2025} and \cite{Agundez2025}, include observations carried out from November 2019 to July 2024, and amount for 1509.2 h of on-source telescope time \citep{Cernicharo2024}. \rev{The $T_A^*$ rms noise level varies between 0.06 mK at 32 GHz and 0.18 mK at 49.5 GHz.}

\subsection{\rev{Observational results}} \label{sec:observation_results}

An inspection of the \textsc{Quijote} data reveals a clean and clear emission line that can be assigned to the 2$_{0,2}$-1$_{0,1}$ rotational transition of c-FA (see middle right panel in Fig.\,\ref{fig:line}). This is the brightest expected emission line in the Q band at the low gas temperature of \mbox{TMC-1}. We also searched for the lines of t-FA and found three emission lines that could be unambiguously assigned to the 2$_{1,2}$-1$_{1,1}$, 2$_{0,2}$-1$_{0,1}$, and 2$_{1,1}$-1$_{1,0}$ rotational transitions of t-FA (see left panels in Fig.\,\ref{fig:line}). Assuming that the emission of t-FA is distributed as a circle with a radius of 40$''$, as other molecules whose emission has been mapped in \mbox{TMC-1} \citep{Cernicharo2023}, a rotation diagram constructed with the three lines of t-FA observed \rev{(see line parameters in Table\,\ref{table:lines})} in the Q band yields a rotational temperature of 7.0\,$\pm$\,1.0 K \rev{shown in Appendix \ref{sec:app:a}}. The column density derived for t-FA is (9.3\,$\pm$\,0.5)\,$\times$\,10$^{11}$ cm$^{-2}$, assuming local thermodynamic equilibrium (LTE) at this rotational temperature. We noticed that for t-FA the 2$_{1,1}$-1$_{1,0}$ and 2$_{1,2}$-1$_{1,1}$ lines are twice less intense than the 2$_{0,2}$-1$_{0,1}$ line, while in the case of c-FA, these two lines are considerably weaker with respect to the 2$_{0,2}$-1$_{0,1}$ line (see bottom and top right panels in Fig.\,\ref{fig:line}). \rev{In fact, these lines are only marginally detected}. The different relative line intensities observed for t-FA and c-FA indicate a distinct excitation probably attributed to the marked difference in the $b$-type dipole moment, 0.2096 D for t-FA \citep{Kuze1982} and 2.71 D for c-FA \citep{Hocking1976}. We find that a rotational temperature of 4.5 K reproduces the relative intensities of the three available lines of c-FA. The column density derived for c-FA is (5.3\,$\pm$\,0.5)\,$\times$\,10$^{10}$ cm$^{-2}$, also assuming an emission size of 40$''$ of radius. Therefore, the abundance of the cis isomer of formic acid in \mbox{TMC-1} is $\sim$\,6\,\% of that of the trans isomer (Trans / Cis ratio of 17.5), in excellent agreement with the values derived in the cold dense clouds B5 \citep{Taquet2017} and L483 \citep{Agundez2019}, and substantially lower than the value of $\sim$\,35\,\% (Trans / Cis ratio of 2.9) derived in the Orion Bar PDR \citep{Cuadrado2016}, where photoswitching seems to be the dominant formation pathway to c-FA.

\section{Chemical models}

\subsection{Model description}

\begin{table}[t]
   \begin{center}
   \caption{Initial physical conditions in our models. n$_{\text{\ce{H2}}}$ is the \ce{H2} density, A$_v$ the extinction coefficient, $\zeta$ the cosmic-ray ionization rate, $T_g$, $T_{\rm d}$ the gas and dust temperature, respectively.}
   \label{tab:phys}
   \begin{tabular}{cc}
   \hline
   Parameter & Value \\
   \hline
   n$_{\text{\ce{H2}}}$   &  1$\times$10$^{4}$ cm$^{-3}$ \\ 
   $A_{\rm v}$ & 10 mag \\
   $\zeta$ & 1.3$\times$10$^{-17}$ s$^{-1}$ \\
   $T_{\rm g}$ & 10 K \\
   $T_{\rm d}$ & 10  K \\
   \hline
   \end{tabular}
   \end{center}
   \end{table}

To interpret the observed t-FA/c-FA ratios, we performed astrochemical modeling using rate equations. Specifically, we employed the three-phase chemical model \textsc{Rokko}, as described in \citet{Furuya2015}. The chemical network was based on a modified version of the \citet{Garrod2013} network, adapted ad hoc to incorporate recent advances in our understanding of formic acid chemistry. All major modifications to the reaction network are enumerated in Appendix \ref{sec:app:additions}. The physical conditions of our chemical model are the standard ones for a dark cloud model \citep{Agundez2013} and are gathered in Table \ref{tab:phys}. Grain chemistry is simulated in the model assuming that all potential reaction events follow a diffusion-reaction competition scheme \citep{Chang2007}, where the most important quantity in this context is the binding energy (BE) of the H atom that we set to 450 K and made equal to the one of \ce{H2} \rev{\citep{2024MolPh.12252100H, 2007ApJ...668..294C}}.  Besides, the binding energy of C atoms is set to 10,000 K to reflect chemisorption \citep{Molpeceres2021c,Potapov2021}. All the other binding energies are extracted directly from the \citet{Garrod2013} reaction network, except for FA and HOCO, where our own derived values are used \citep{Molpeceres2022c, Molpeceres2023}.\footnote{E$_{\textrm bin}$(c-HOCO)=6190 K, E$_{\textrm bin}$(t-HOCO)=7302 K, E$_{\textrm bin}$(c-HCOOH)=6227 K, E$_{\textrm bin}$(t-HCOOH)=6170 K } The \ce{H2} surface coverage is treated as an adsorption-desorption equilibrium \citep{Furuya2019,Furuya2024}. The diffusion energy of a given adsorbate is defined as 0.4 its binding energy \rev{\citep{2017SSRv..212....1C}}, except for the case of CO, where the experimentally derived value of $\sim$ 0.25 is used \citep{Furuya2022}\footnote{\rev{Noting that a value of 0.4 for the ratio between diffusion and desorption is not universal, the chemistry of FA is mostly dominated by CO diffusion, for which we set the experimental constrained value \citep{Furuya2022}}} We neglect mantle chemistry in our model, both thermal and non-thermal \citep{Jin2020} allowing only thermal surface chemistry in the four topmost layers of the dust particle, that is modeled assuming 0.1 $\mu$m radius particles. 

To properly reproduce gas-phase abundances of molecules formed on the surface of dust grains we included in the model a series of non-thermal desorption mechanisms into the model, namely photodesorption, both from the interstellar (attenuated) radiation field and the secondary UV field, with a yield of 1$\times$10$^{-3}$. Second, we included stochastic heating of cosmic dust caused by cosmic-ray grain interaction with a temperature of 70 K and a time of 10 $\mu$s per collision event \citep{hasegawa_three-phase_1993}. The most important non-thermal effect included in our simulations is chemical desorption (CD), for which we consider the \citet{garrod_non-thermal_2007} formulation with an $a$ coefficient of 0.01. \rev{In the \citet{garrod_non-thermal_2007} $a$ is the ratio of the bond frequency of the adsorbate-surface normal mode over the rest of vibrational modes that are used for relaxation, always assuming reaction energy equipartition}.  The astrochemical community has still not reached a consensus around the feasability of CD for relatively large organic molecules like FA. However, our recent simulations \citep{Molpeceres2023b} showed that CD probability is binding site dependent, hence supporting the notion that it should remain a possibility for most adsorbates. New to this work is the split of chemical desorption into the different product channels of FA:

\begin{align}
   \ce{t-HOCO(i) + H(i) &-> c-HCOOH(g)} \label{eq:iud1} \\
   \ce{ &-> t-HCOOH(g)}, \label{eq:iud2}
\end{align}
where (i) and (g) indicate ``ice'' and ``gas''. The logic behind this implementation is that, when a molecule desorbs through CD, it does with a fraction of the reaction energy that could not be dissipated. Because the ice partly dissipates the energy \citep{Pantaleone2021, Ferrero2023, Molpeceres2023b} dissociation is an unlikely outcome, and therefore isomerization to exothermic channels is possible. This mechanism is labeled in this work as isomerization-upon-desorption (IUD) and must be dependent on the fraction of energy remaining on the molecule after desorption. Because IUD is an effect that warrants a more in-depth investigation, we performed a sensitivity analysis on the impact of varying c/t ratios in this reaction (Section \ref{sec:results}), varying $a$ between the two reaction channels (they both sum the 0.01 value used for the rest of the molecules) and making it the free parameter of the model. While an IUD probability can also be defined from the reaction \ce{c-HOCO(i) + H(i)}, in this case the process points in the endothermic direction and thus contributes less significantly. Nevertheless, the effect of this channel is implicitly accounted for in the sensitivity analysis used to constrain the best-fitting model using reactions \ref{eq:iud1} and \ref{eq:iud2}, as discussed in Section~\ref{sec:results}.

\begin{table}
   \caption{ Initial elemental abundances with respect to H nuclei. Taken from \citet{Aikawa1999}.}
   \label{tab:initial_abundances}
   \centering
   \begin{tabular}{cc}
   \hline
   Element  & Initial Abundance  \\
   \hline
   \ce{H2}  & 0.5    \\
   \ce{He}  & 9.8$\times10^{-2}$ \\ 
   \ce{N}   & 2.5$\times10^{-5}$ \\
   \ce{O}   & 1.8$\times10^{-4}$ \\
   \ce{C+}  & 7.9$\times10^{-5}$ \\
   \ce{S+}  & 9.1$\times10^{-8}$ \\
   \ce{Si+} & 9.7$\times10^{-9}$ \\
   \ce{Fe+} & 2.7$\times10^{-9}$ \\
   \ce{Na+} & 2.3$\times10^{-9}$ \\
   \ce{Mg+} & 1.1$\times10^{-8}$ \\
   \ce{Cl+} & 2.2$\times10^{-10}$ \\
   \ce{P+}  & 1.0$\times10^{-9}$ \\
   \hline
   \end{tabular}
   \tablefoot{\rev{We note that \ce{S+} abundances are considered significantly higher in, for example, TMC-1-CP \citep{fuente_gems_2023}. However, the models for FA are rather insensitive to sulfur abundance, and we keep \citet{Aikawa1999} values for consistency.}}
\end{table}

As mentioned above, the reaction network used in this work is a significantly modified version of the \citet{Garrod2013} one. Apart from \ce{H2} all the remaining elements are set in its atomic form at the beginning of the simulations, either neutral or ionic depending on their ionization potential (Table \ref{tab:initial_abundances}). \rev{We also tested different sets of initial abundances, including the higher metallicity values proposed by \citet{wakelam_efficiency_2021}. These tests showed a noticeable effect on the overall abundances with minor importance on the c/t ratios. The low-metallicity values are motivated by the ratio of microcanonical rate constants for isomerization discussed in Section \ref{sec:results}. It should be stressed, however, that our results depend on the IUD introduced above, which remains a free parameter of the model. A more extensive exploration of this parameter space would be warranted if the model were to be extended, although for our present purposes the current choice is sufficient.}  

The main change in the chemical reaction network pertaining this article is the duplication of all HOCO and HCOOH entries in the network into c-HOCO, t-HOCO, c-HCOOH and t-HCOOH, the inclusion of the whole FA/\ce{CO2} reaction network compiled from a series of works \citep{GarciadelaConcepcion2022,Molpeceres2022c, GarciadelaConcepcion2023,Molpeceres2023, Molpeceres2025} expanded with several gas-phase destruction reactions, and the abovementioned inclusion of IUD. Finally, along with IUD, we also implemented an additional type of reaction, unimolecular isomerization in the gas phase to interconvert trans/cis isomers, particularly FA and HOCO:

\begin{equation}
   \ce{t-HCOOH(g) <=> c-HCOOH(g)}
\end{equation}
or:
\begin{align}
   \ce{t-HCOOH(g) &-> c-HCOOH(g)} \label{eq:iso:tFAcFA} \\
   \ce{c-HCOOH(g) &-> t-HCOOH(g)} \label{eq:iso:cFAtFA}. 
\end{align}
The rate constant for this process is simply expressed as $k_{\rm iso}$=$\alpha$ where $\alpha$ is a value (not interpolated)\footnote{The reason for not interpolating the rate constants is the dramatic dependence of the rate constant with temperature at low temperatures, we prefer to extract the exact values provided in \citet{GarciadelaConcepcion2021}} extracted from Table 1 of \citet{GarciadelaConcepcion2022}. An enumeration of all the changes in the reaction network is collated in Appendix \ref{sec:app:additions}, with the most important reactions being (some of them are already mentioned):

\rev{
\begin{alignat}{2}
   \ce{CO(i) + OH(i) &-> t-HOCO(i)}             & E_\mathrm{a} = 100~\text{K} \label{eq:array:1} \\
   \ce{CO(i) + OH(i) &-> c-HOCO(i)}             & E_\mathrm{a} = 100~\text{K} \label{eq:array:2} \\
   \ce{c-HOCO(i) + H(i) &-> t-HCOOH(i)}         & E_\mathrm{a} = 0~\text{K}     \label{eq:array:3} \\
   \ce{c-HOCO(i) + H(i) &-> t-HCOOH(g)}         & E_\mathrm{a} = 0~\text{K}     \label{eq:array:4} \\
   \ce{t-HOCO(i) + H(i) &-> c-HCOOH(i)}         & E_\mathrm{a} = 0~\text{K}   \label{eq:array:5} \\
   \ce{t-HOCO(i) + H(i) &-> c-HCOOH(g)}         & E_\mathrm{a} = 0~\text{K}   \label{eq:array:6} \\
   \ce{t-HOCO(i) + H(i) &-> t-HCOOH(g)}         & E_\mathrm{a} = 0~\text{K}   \label{eq:array:7} \\
   \ce{c-HCOOH(i) + H(i) &-> t-HOCO(i) + H2(i)} & E_\mathrm{a} = 3240~\text{K}  \label{eq:array:8} \\
   \ce{t-HCOOH(i) + H(i) &-> c-HOCO(i) + H2(i)} & E_\mathrm{a} = 4830~\text{K}  \label{eq:array:9} \\
   \ce{t-HCOOH(g) &<=> c-HCOOH(g)}              & \text{See footnote 4}  \label{eq:array:10} \\
   \ce{c-HOCO(i) + H(i) &-> CO2(i) + H2(i)}              & E_\mathrm{a} = 301~\text{K}   \label{eq:array:11} \\
   \ce{t-HOCO(i) + H(i) &-> CO2(i) + H2(i)}              & E_\mathrm{a} = 7340~\text{K}   \label{eq:array:12} \\
   \ce{c-HOCO(i) + H(i) &-> CO(i) + H2O(i)}              & E_\mathrm{a} = 4120~\text{K}   \label{eq:array:13} \\
   \ce{t-HOCO(i) + H(i) &-> CO(i) + H2O(i)}              & E_\mathrm{a} = 0~\text{K}   \label{eq:array:14}
\end{alignat}
}

The HOCO formation reactions (reactions \ref{eq:array:1} and \ref{eq:array:2}) are described in \citet{Molpeceres2023}\footnote{Because these reactions are essentially almost barrierless and non-thermal effects are likely at play, we model it assuming a barrier of merely 100 K \citep{Molpeceres2023}.}. The HOCO hydrogenation reactions (Reactions \ref{eq:array:3}--\ref{eq:array:7}; and competition channels \ref{eq:array:11}--\ref{eq:array:14}) are modeled after \citet{Molpeceres2025}. The \ce{H-abstraction} reactions (reactions \ref{eq:array:8}-\ref{eq:array:9}), for which the rate constants are fitted to match accurate instanton calculations are taken from \citet{Molpeceres2022c}. The trans-cis equilibrium (i.e. after desorption), with forward and backward rate constants, is taken from \citet{GarciadelaConcepcion2022}, as indicated above. Lastly, the value of the IUD reaction (\ref{eq:array:6} and \ref{eq:array:7}) are obtained from the sensitivity study shown in the next section.

\begin{figure}
   \centering
   \includegraphics[width=\columnwidth]{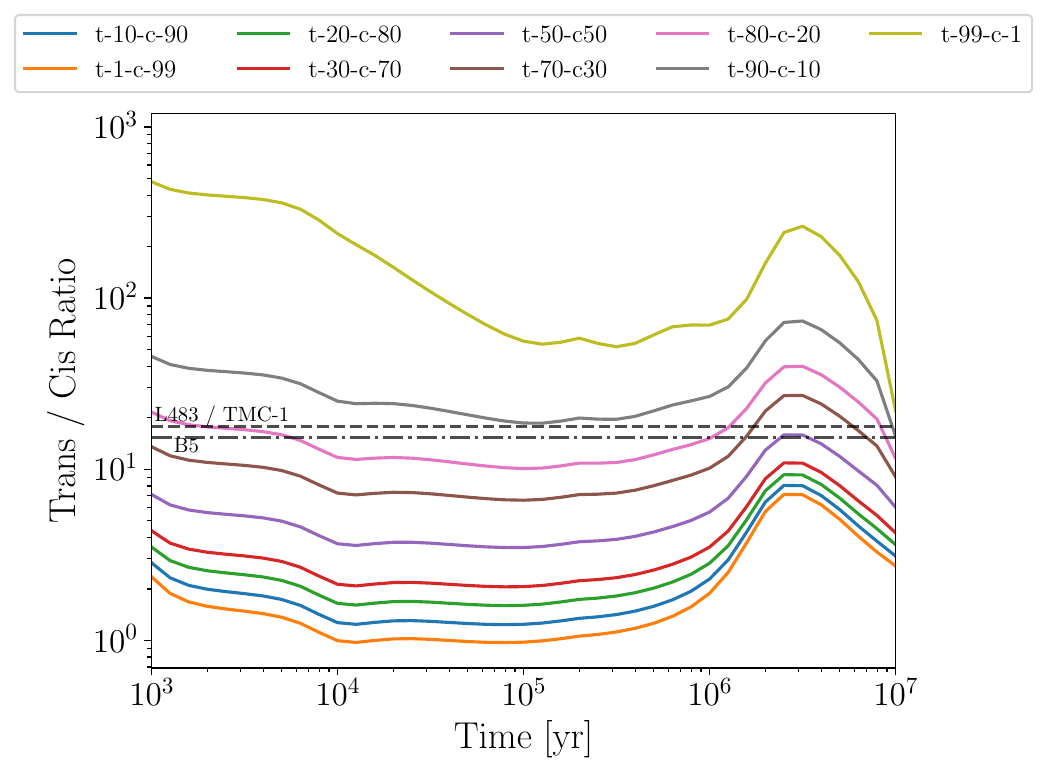}
      \caption{t-FA / c-FA ratio over time as a function of the isomerization upon desorption (IUD) ratio. In the legend t and c indicate the isomeric form and the numbers the percentage of desorption of each of them in the \ce{t-HOCO + H -> (t/c)-HCOOH} reaction. Horizontal \rev{lines} represent the gas-phase observational abundances of both isomers in the Barnard-5 molecular cloud \citep{Taquet2017}, L483 \citep{Agundez2019} and TMC-1 (This work).
              }
         \label{fig:ratio}
   \end{figure}

\begin{figure}
      \centering
      \includegraphics[width=0.9\columnwidth]{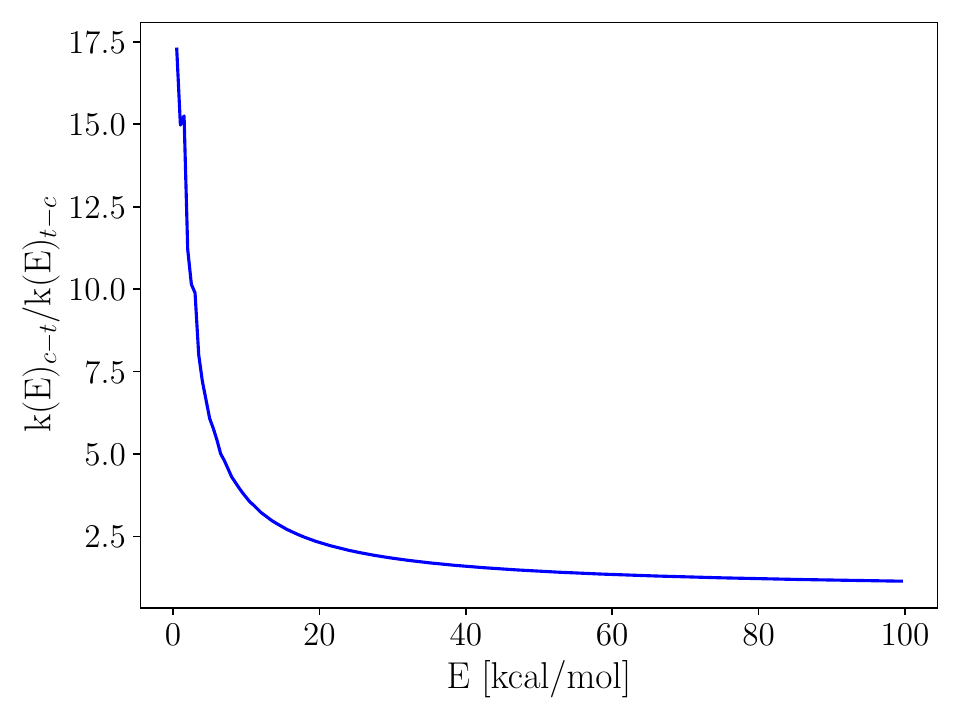}
         \caption{Microcanonical rate constant ratio for c-FA and t-FA isomerization upon desorption. The limit on the x-axis is determined from the reaction energy for the HOCO radical hydrogenation on ices \citep{Molpeceres2025} in between 90-105 kcal mol$^{-1}$ depending on factors as the HOCO isomer being hydrogenated or the ice binding site. The energy of c-FA is set as the origin of energies.
                 }
            \label{fig:ke}
\end{figure}

\subsection{Results} \label{sec:results}

To determine the optimal model for the observed t/c-FA ratio, we performed a sensitivity analysis of the model on the probability of IUD.\footnote{The probability of IUD is referred with respect the most stable molecule of the isomer pair, in this case t-FA. Hence a 90\% IUD indicates that the isomeric imbalance is attributed to only a single reaction of all the possible reaction channels, in this case reactions \ref{eq:iud1} and \ref{eq:iud2}} In essence, this analysis allows to constraint the branching ratio of reactions \ref{eq:iud1} and \ref{eq:iud2} (and \ref{eq:array:6} and \ref{eq:array:7}). The results of this analysis are shown in Figure \ref{fig:ratio}. The grid of models was run with IUD probabilities ranging from 1\% to 99\% for the formation of t-FA and c-FA. An inspection of Figure \ref{fig:ratio} reveals that an IUD probability between 70\% and 90\% is necessary to reproduce the observed t-FA/c-FA ratios in TMC-1, L483, and Barnard-5, depending on the considered timescales. Among these, the best-performing model, selected as the one matching the characteristic timescale of TMC-1 \citep[in the order of 10$^{5}$ yr][\rev{although in our simulations the abundances of B5 are better described than those of TMC-1}]{Pratap1997} corresponds to an IUD probability of 90\%, indicating that chemical desorption followed by instantaneous isomerization plays a crucial role in determining the correct t/c-FA ratio and likely influences other interstellar molecules. We adopt the 90\% IUD model for the subsequent discussions.

The selection of a high IUD probability can be rationalized by examining the microphysics of chemical desorption. After a chemical desorption event, the isomerization process resembles a microcanonical reaction in the gas phase where the energy available for the reaction is not known, but should be significantly lower than the total reaction energy \citep{Molpeceres2023b}, reducing the likelihood of dissociation. The ratio of the microcanonical rate constants for the \ce{c-FA <=> t-FA} isomerization, computed using DFT calculations ($\omega$B97M-D4/def2-TZVPPD; \cite{mardirossian_2016,Caldeweyher2019,najibi_dft_2020,rappoport2010a,weigend2005a}) and RRKM theory including quantum tunneling, is shown in Figure \ref{fig:ke}. \rev{We use \textsc{Orca6.0.0} and \textsc{Mess} for these calculations \citep{Neese2012,Neese2020,neese_software_2022,Georgievskii2013}.} These rate constants resemble the ones shown in \citet{GarciadelaConcepcion2021}, but in the microcanonical ensemble, that is the one applicable just after a chemical desorption. Without a substrate to interact with, the excess energy can drive back-and-forth isomerizations. The microcanonical rate constant ratio for c-FA to t-FA ranges from approximately 17.5 to 2.0, depending on the residual energy in the molecule after desorption. This indicates that during desorption events, the isomerization of c-FA to t-FA is strongly favored. This aligns well with the results in Figure \ref{fig:ratio}, where models incorporating significant IUD probabilities yield better agreement with observations. Furthermore, since not all desorption events impart the same amount of energy to the desorbing molecule, the average ratio of rate constants will be higher than 2.0. It is improbable that the full reaction energy (approximately 100 kcal mol$^{-1}$ for the HOCO hydrogenation reaction; \citealt{Molpeceres2025}) remains in the molecule after desorption, and if so, the molecule would probably simply dissociate as mentioned above. This strongly supports a high IUD probability, as indicated by our model. However, the exact energy retained by the desorbed molecule is not well constrained and depends on how much energy the dust matrix absorbs during the non-thermal desorption event. It is worth noting that additional non-thermal desorption mechanisms may also influence the t-FA/c-FA ratio. The adoption of a 90\% IUD probability might inadvertedly account for other \rev{isomerization processes after desorption}, such as cosmic-ray sputtering \citep{Dartois2021} or other non-thermal mechanisms like desorption by shocks, which could contribute to the observed ratios alongside chemical desorption.

\begin{figure*}
   \centering
   \includegraphics[width=0.33\linewidth]{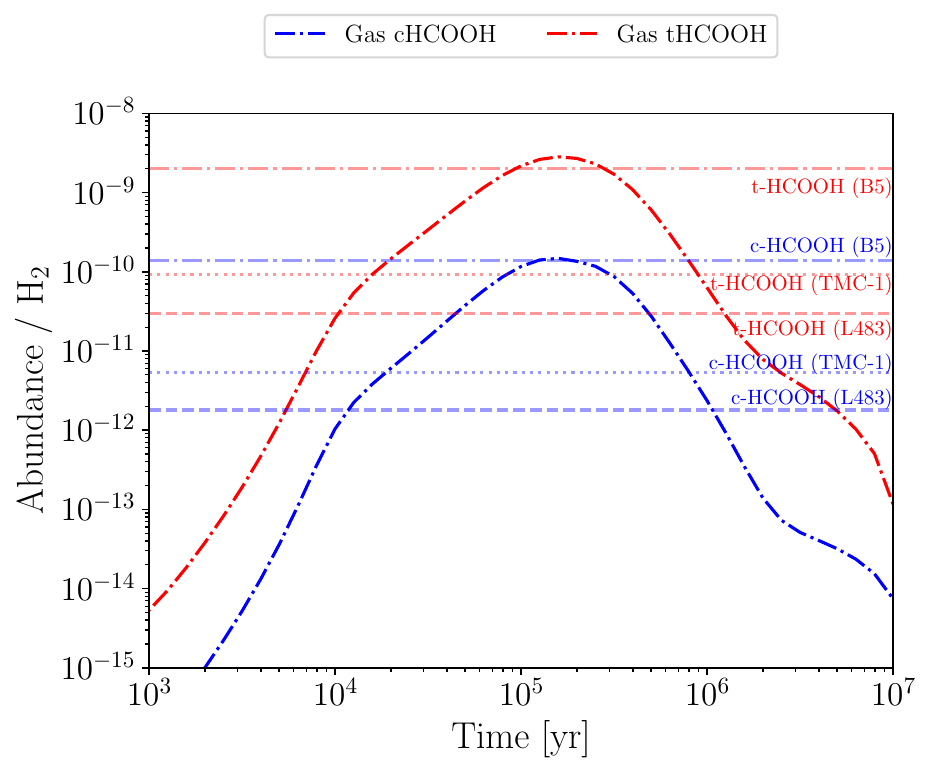}
   \hfill
   \includegraphics[width=0.33\linewidth]{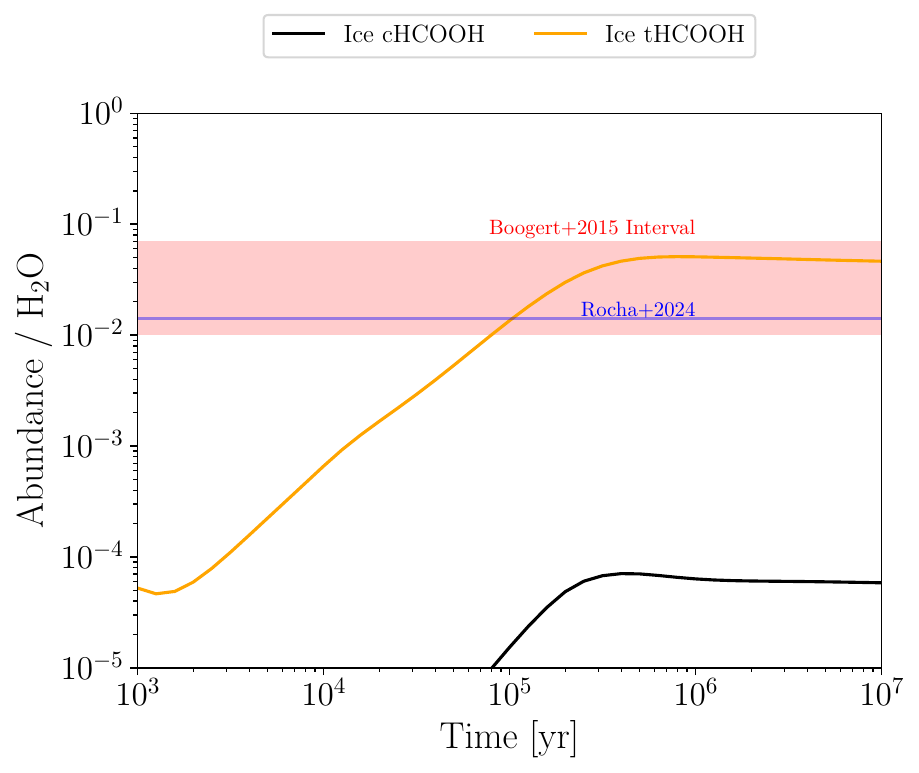}
   \hfill
   \includegraphics[width=0.33\linewidth]{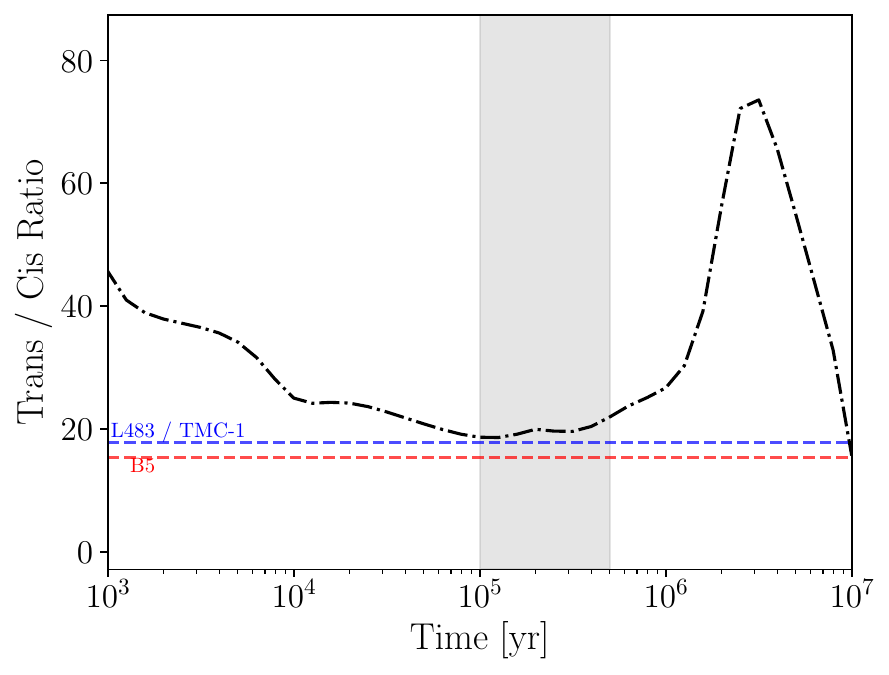}
      \caption{\rev{Left panel - Chemical model using the best performing model for c-FA and t-FA in the gas. Horizontal bands represent the range of observational abundances with respect to \ce{H2} of both isomers in the three clouds where they have been detected, the Barnard-5 molecular cloud \citep{Taquet2017}, L483 \citep{Agundez2019} and TMC-1 (This work). Middle panel - Equivalent to above but for ice abundances (comparing with \ce{H2O}) and comparing with the upper and lower bounds provided in \citet{Boogert2015} and with the well constrained narrow band of \citet{Rocha2024}. We note that \ce{HCOOH} and \ce{HCOO-} are summed in establishing the observational abundances  Right - t-FA/c-FA gas phase ratios for the best performing model in linear scale. The gray area represents an approximated characteristic timescale of TMC-1 \citep{Pratap1997}.
      }}
         \label{fig:abundances}
   \end{figure*}

After setting the free parameter of our model, the IUD probability, we run a chemical model for $T_{\rm g} = T_{\rm d} = 10$ K, equivalent to those where these c-FA have been detected \citep{Taquet2017,Agundez2019}, which serves as the main reference for discussing FA isomerism in dark clouds. The results of this model are presented in Figure \ref{fig:abundances}, where in the bottom panel we copy the results from Figure \ref{fig:ratio} in a linear scale. A visual inspection reveals a good match between our gas-phase values and the observational data from Barnard 5 (B5) \citep{Taquet2017}, L483 \citep{Agundez2019} and TMC-1. Given the strong agreement at $1$–$5\times10^{5}$ yr in absolute abundances and the predicted t-FA/c-FA ratio (vide infra), we consider that differences in the absolute abundances of t-FA and c-FA in different sources to the physical conditions of the modeled objects rather than to limitations in the chemical network. \rev{Simultaneously reproducing the chemical abundances of three objects with different physical conditions and evolutionary stages is not feasible, particularly when using a pseudo–time-dependent single-point physical model. Nevertheless, the level of agreement achieved is between excellent and acceptable across the three sources, suggesting that the differences arise primarily from the physical conditions of each source rather than from missing chemical reactions needed to account for FA formation and its isomerism. } Our models also show excellent agreement with total FA ice abundances in young stellar objects, predicting a 2–7\% abundance relative to \ce{H2O} ice, consistent with the ranges reported by \citet{Boogert2015} and recently by \citet{Rocha2024}. It is important to note that our comparison with observational data includes both neutral FA and the \ce{HCOO-} ion, an immediate proxy that is not explicitly included in our model. A limitation of our model is that FA and \ce{CO2} ice (not shown in Figure \ref{fig:abundances}) are predicted to have similar abundances, whereas observational constraints suggest a \ce{CO2}/FA ice ratio of 2–3 in favor of \ce{CO2} \citep{Boogert2015, McClure2023}. While this discrepancy does not affect the conclusions of this study, we discuss potential ways to reconcile our models with observations in Appendix \ref{sec:app:co2}. The agreement portrayed in both panels Figure \ref{fig:abundances} is explained by the combination of reactions \ref{eq:array:1}--\ref{eq:array:14}, but it is fair to stress out that this chain of reactions can \emph{only} initiate if Reactions \ref{eq:array:1} and \ref{eq:array:2} proceed as shown in \citet{Molpeceres2023} and recently in \citet{Ishibashi2024}. This is consistent with Kinetic Montecarlo simulations that suggest the importance of additional \ce{CO2} formation reactions \citep{jimenez-serra_modelling_2025}

After Reactions \ref{eq:array:1} and \ref{eq:array:2} the isomer imbalance of FA is caused by the H-addition/\ce{H}-abstraction cycle shown in \citet{Molpeceres2022c,Molpeceres2025}. Briefly, atomic H partakes easily on c-FA \ce{H}-abstraction (Reaction \ref{eq:array:8}) whereas the same reaction in t-FA is very slow (Reaction \ref{eq:array:9}), creating an isomeric imbalance of several orders of magnitude for the abundances on the ice, Figure \ref{fig:abundances} \rev{middle panel}. However, a more detailed investigation of Figure \ref{fig:abundances} shows that the t-FA / c-FA ratio is markedly lower in the gas (\rev{left panel}), owing to the increased CD coming from a H-addition/\ce{H}-abstraction loop that increases the reactive events and therefore the total CD rate for c-FA. This cycle is not present for t-FA because, as we said above, Reaction \ref{eq:array:9} is slow, hence the postulation of the IUD mechanism, that we discussed in detail above. Once in the gas, c/t-FA interconversion is, at 10 K, strictly out of equilibrium and the chemistry of each isomer can be controlled either by destruction reactions, e.g. with \ce{C+}, \ce{HCO+}, \ce{H3+} where the isomer with larger dipole moment is more favorably destroyed \citep[\rev{and effect labelled the relative dipole principle,}][]{Shingledecker2020}, or by unimolecular isomerization. At 10 K, unimolecular isomerizations rates are negligible for both isomers, and therefore destruction reactions modulate the isomeric balance that comes from the desorption of the grain.

\begin{figure}
   \centering
   \includegraphics[width=\columnwidth]{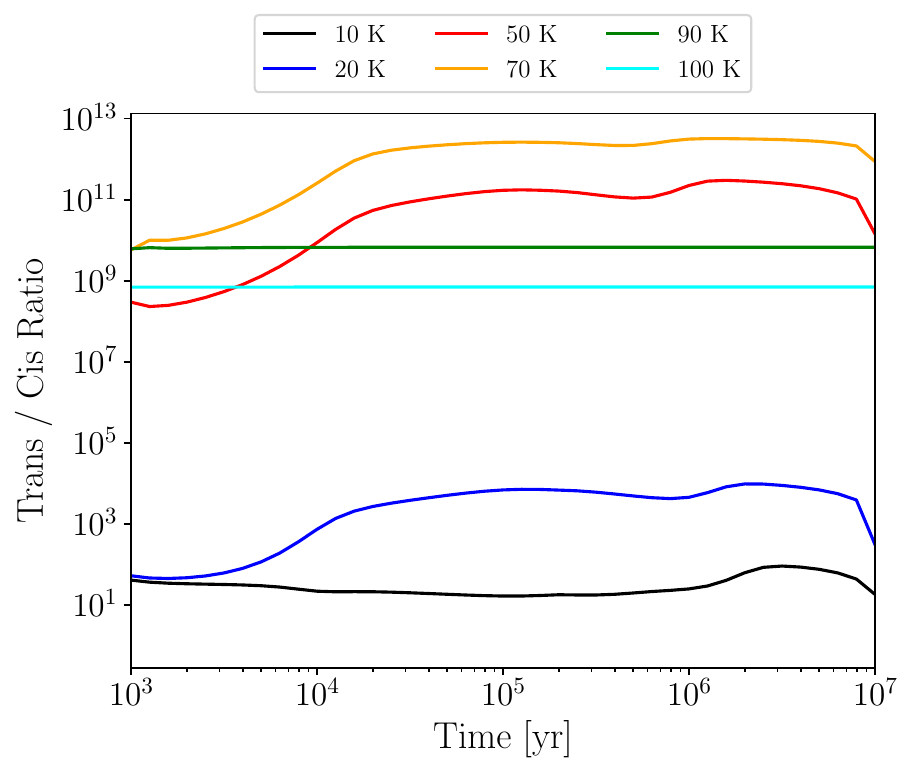}
      \caption{Gas phase t-FA / c-FA for different $T_{\rm g}$. $T_{\rm d}$ is kept constant (10 K). Horizontal lines indicate that chemical equilibrium is reached. The equilibrium constants are, according to \citet{GarciadelaConcepcion2022}, K$_{\rm t/c}$(10 K)=2.1(88), K$_{\rm t/c}$(20 K)=1.5(44), K$_{\rm t/c}$(50 K)=4.7(17), K$_{\rm t/c}$(70 K)=4.3(12), K$_{\rm t/c}$(90 K)=6.7(9), K$_{\rm t/c}$(100 K)=7.0(8). A(B) is equivalent to A$\times$10$^{B}$. }
         \label{fig:ratioT}
   \end{figure}

In order to investigate the competition of unimolecular isomerization (Reaction \ref{eq:array:10}) with destruction reactions in the gas for c/t-FA ratios after it is desorbed from the grains we carried out additional numerical experiments simulating a series of ficticius dark clouds where we vary T$_{\rm g}$ and assume a thermal decoupling of gas and grains (T$_{\rm d}$ = 10 K). The results are shown in Figure \ref{fig:ratioT}. The first conclusion extracted from the figure is that only at 10 K, the ratio is sufficiently low (2$\times$10$^{1}$) to ensure easy c-FA detectability, despite the extremely high equilibrium t-FA / c-FA equilibrium constant (2.1$\times$10$^{88}$). At 20 K, this ratio is above 10$^{3}$ which is low, but detection of c-FA might still be possible. At much higher temperatures, the ratio predicted by the models starts to converge to the thermodynamic one. It is important to note, however, that given the dramatic dependence of the thermal isomerization rate constant with temperature, and the inherent limitations of the theoretical predictions for these rate constants, the value of 20 K is subjected to uncertainties. At 50 and 70 K, FA destruction reactions including the sequential acid base mechanism shown in \citet{GarciadelaConcepcion2023} preclude reaching formal thermodynamic equilibrium because their reaction rate surpasses the one of Reaction \ref{eq:array:10} (from right to left). This threshold is surpassed above 90 K, where it is safe to consider thermodynamic equilibrium, that is characterized by a totally horizontal line in the time evolution of the t-FA / c-FA ratios (Figure \ref{fig:ratioT}). The exact temperature at which thermodynamic equilibrium is reached is difficult to pinpoint, as it strongly depends on the completeness of the chemical network. For example, if currently unexplored reactions in the FA network, such as \ce{HOCO+ + H2}, play a significant role, the upper limit of the equilibrium temperature range may shift accordingly. At lower temperatures, the contribution from grain-surface processes clearly dominates. To sum up, the immediate conclusion of this set of models is that c-HCOOH in dark clouds is a consequence of out of equilibrium conditions and that its temperature range of existence is narrow. Our models predict a dramatic decrease of c-HCOOH abundance under the detectability limits at temperatures above 50 K, explaining, for example, its absence in hot cores. In environments that are more energetic than cold dark clouds but not as extreme as hot cores, such as the G+0.693–0.027 molecular cloud, where shocks and elevated cosmic-ray ionization rates are present, formic acid may be gradually released from dust grains into the gas phase. This could account for detections like our recent observation of \ce{t-HCOOCH3} \citep{sanz-novo_conformational_2025}. However, if the tentative detection of c-FA in this cloud \citep{SanzNovo2023} is confirmed, the resulting t-FA/c-FA ratios would be at least an order of magnitude higher than those typically found in dark clouds.

The stringent conditions for the detectability of c-FA make this molecule a viable cloud temperature tracer, although such a role can only be effectively exploited with very precise rate constants for isomerization, likely exceeding our current theoretical frameworks. Finally, we remind that the rationalizations made in this work are only applicable in the absence of a strong interstellar UV field, as it has been reported that near IR pumping is capable of explaining the presence of c-FA in photodissociation regions \citep{Cuadrado2016}


\section{Conclusions: formic acid isomerism in dark clouds}
The picture that emerges from our detection of c-FA in TMC-1 and subsequent chemical modeling is as follows. The total abundance of formic acid can only be explained by considering its formation via the HOCO intermediate through the \ce{CO + OH} reaction. When this pathway is included, FA ice abundances align well with those reported in ice observations. Introducing the isomeric dimension reveals that most FA on ice exists as t-FA, owing to its greater resistance to hydrogenation compared to c-FA.  The lower resistance of c-FA on ice to hydrogenation triggers an H-abstraction and H-addition cycle that promotes chemical desorption, enriching the cold gas with FA. Crucially, the inclusion of the "isomerization upon desorption" (IUD) mechanism is key to matching observations. This mechanism suggests that any excess energy released during a non-thermal desorption event (e.g., chemical desorption) can subsequently drive isomerization. By incorporating this mechanism, our model successfully reproduces the observed abundances of t-FA and c-FA in dark clouds, as well as their relative ratios. A detailed analysis of the temperature dependence of these ratios reveals that, in the gas phase, c-FA abundance declines to undetectable levels at temperatures above 50 K and possibly lower. This explains the absence of c-FA in warmer astronomical environments, such as hot cores. The comprehensive rationalization of FA's isomeric chemistry provides a robust framework for understanding the detectability of high-energy isomers in the ISM and paves the way for applying this approach to other molecular species.

Finally, this work underscores the importance of non-thermal, energetic mechanisms in shaping the molecular and isomeric inventory of cold dark clouds. The interplay between these mechanisms and thermal processes is critical for understanding the chemical evolution of the ISM. Our ad-hoc addition of a non-thermal (microcanonical) isomerization mechanism to the chemical network highlights the need for further investigation, which we plan to address in future studies.

\begin{acknowledgements}
   We thank Dr. Kenji Furuya for the newest release of the \textsc{Rokko} code. G.M acknowledges the support of the grant RYC2022-035442-I funded by MCIU/AEI/10.130\
   39/501100011033 and ESF+. G.M. also received support from project 20245AT016 (Proyectos Intramurales CSIC). We acknowledge the computational resources provided by the DRAGO computer cluster managed by SGAI-CSIC, and the Galician Supercomputing Center (CESGA). The supercomputer FinisTerrae III and its permanent data storage system have been funded by the Spanish Ministry of Science and Innovation, the Galician Government and the European Regional Development Fund (ERDF). This publication is part of the project PID2023-147545NB-I00 funded by MICIU/AEI/10.13039/501100011033. M.S.-N acknowledges a Juan de la Cierva Postdoctoral Fellow project JDC2022-048934-I, funded by MICIU/AEI/10.13039/501100011033 and by the European Union "Next GenerationEU/PRTR". V.M.R. acknowledges support from the grant PID2022-136814NB-I00 by the Spanish Ministry of Science, Innovation and Universities/State Agency of Research MICIU/AEI/10.13039/501100011033 and by ERDF, UE; the grant RYC2020-029387-I funded by MICIU/AEI/10.13039/501100011033 and by "ESF, Investing in your future", and from the Consejo Superior de Investigaciones Cient{\'i}ficas (CSIC) and the Centro de Astrobiolog{\'i}a (CAB) through the project 20225AT015 (Proyectos intramurales especiales del CSIC); and from the grant CNS2023-144464 funded by MICIU/AEI/10.13039/501100011033 and by “European Union NextGenerationEU/PRTR”. I.J-.S acknowledges funding from the ERC Consolidator grant OPENS (project number 101125858) funded by the European Union, and from grant PID2022-136814NB-I00 funded by the Spanish Ministry of Science, Innovation and Universities/State Agency of Research MICIU/AEI/ 10.13039/501100011033 and by “ERDF/EU”. JGdlC also acknowledge European Funds of Regional Development, and the Autonomous Government of Extremadura (Grant GR24020).

\end{acknowledgements}

%
%

\appendix

\section{Rotation diagram for t-FA} \label{sec:app:a}

\begin{figure}
\centering
\includegraphics[width=0.9\columnwidth]{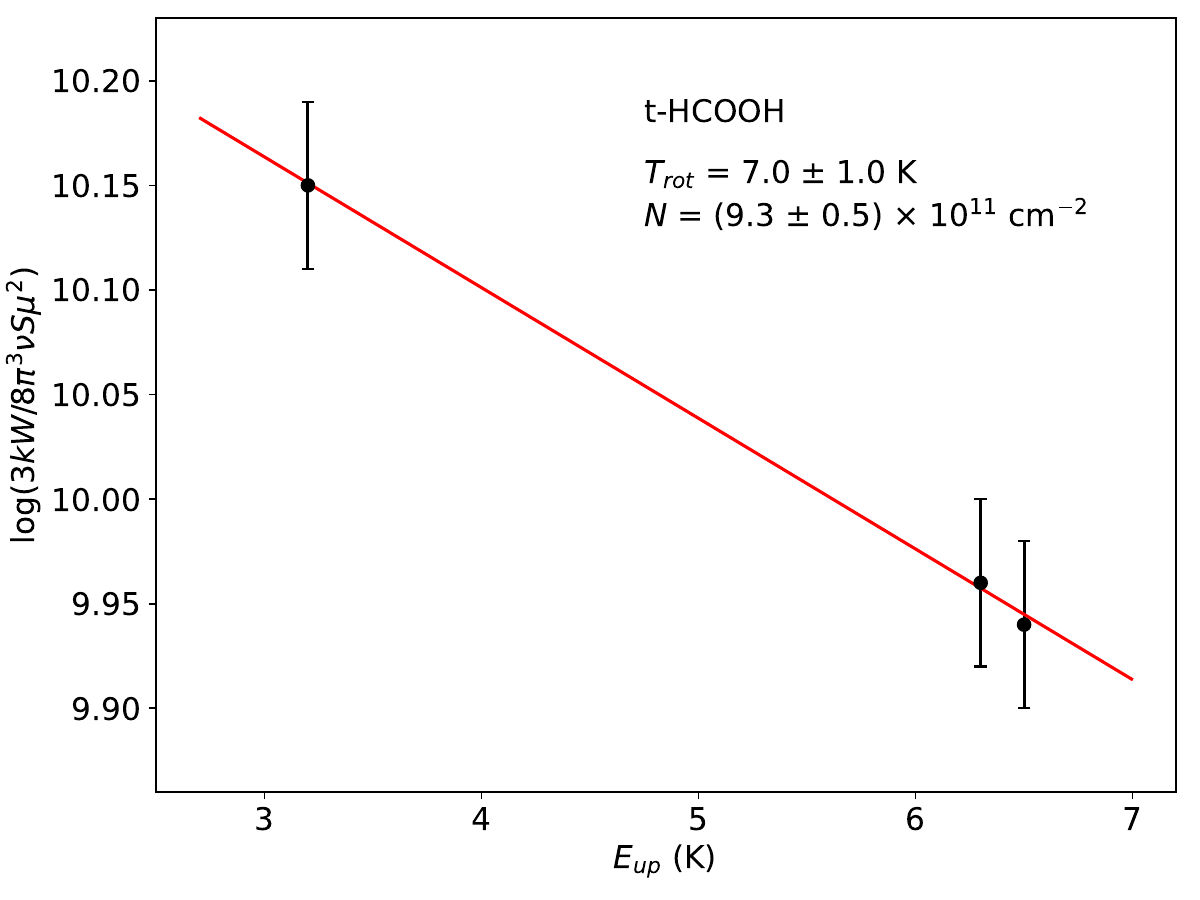}
\caption{Rotation diagram for t-HCOOH in TMC-1.}
\label{fig:rotation_diagram}
\end{figure}

\section{Summary of additions to the reaction network} \label{sec:app:additions}

A summary of the main modifications of the \citet{Garrod2013} reaction network used in this work is enumerated below. The full reaction network can be retrieved at \url{https://zenodo.org/records/17179209}. In the reaction network provided \texttt{COOH} and \texttt{HCOOH} are the old isomer non-inclusive HOCO radical and FA for whom all reactions are deactivated. By contrast, the isomer inclusive species are \texttt{c-COOH}, \texttt{t-COOH}, \texttt{c-HCOOH} and \texttt{t-HCOOH}.

\begin{itemize}
   \item Substitution of conventional \ce{CO2} chemistry through the \ce{CO + OH -> CO2 + H} reaction by the newer \ce{CO + OH -> c/t-HOCO} followed by hydrogenation, \ce{HOCO + H -> CO2 + H2}, \ce{HOCO + H -> HCOOH} and \ce{HOCO + H -> CO + H2O} \citet{Molpeceres2023,Ishibashi2024,Molpeceres2025}. 
   \item Splitting of HOCO and HCOOH chemistry into their isomeric forms c-HOCO, t-HOCO, c-HCOOH, t-HCOOH, with the main formation (see below) and destruction reactions (\ce{H3+}, \ce{HCO+}, \ce{C+}) made distinct for each isomer.
   \item Inclusion of carbon atom chemisorption \cite{Molpeceres2021c, Potapov2021, Tsuge2023}
   \item Destruction reactions of HCOOH with \ce{HCO+} and sequential base reaction with \ce{NH3} following \citet{GarciadelaConcepcion2023}.
   \item Updated values of ion-molecule destruction reactions of c-HOCO, t-HOCO, c-HCOOH, t-HCOOH with \ce{C+}, \ce{H3+} and \ce{HCO+}. Values for the rate constant were obtained from the Su-Chesnavich formula \citep{Su1982} and fitted to a 3-term Arrhenius expression. Dipole moments and polarizabilities are taken from \citet{GarciadelaConcepcion2023} for HCOOH and our own quantum chemical calculations (B2PLYP-D3BJ/aug-cc-pVTZ) for HOCO.
   \item Addition of the isomerization-upon-desorption mechanism for \ce{t-HOCO + H} as indicated in the main text.
   \item Inclusion of ion-grain recombination reactions to all cations in the reaction network.
\end{itemize}

\section{Improving \ce{CO2} ice formation} \label{sec:app:co2}

\begin{figure}
   \centering
   \includegraphics[width=\columnwidth]{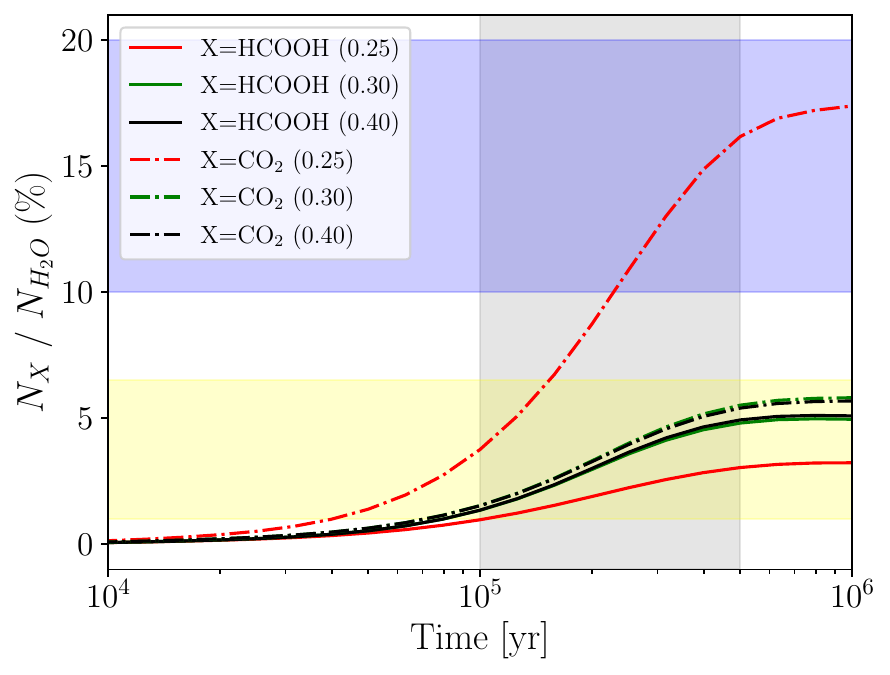}
      \caption{Sensitivity analysis of our model to the oxygen atom diffusion energy for \ce{CO2} ice (dashed lines) and FA (suming c-t abundances). Values are represented with respect to \ce{H2O} column density- The difussion $E_{\rm diff}$ energy is defined from the $E_{\rm diff}/E_{\rm bin}$ parameter in our models. $E_{\rm diff}/E_{\rm bin}$ is set to 0.25, 0.30 and 0.40 in those models. The pale blue zone represents the observationally constrained abundances of \ce{CO2} \citep{Boogert2015,McClure2023} and the pale yellow zone represent the same for FA \citep{Boogert2015,Rocha2024}. We assume that FA and \ce{HCOO-} abundances sum.}
         \label{fig:co2}
   \end{figure}

We mention in the main text that while our model predicts FA abundances and in particular trans-cis ratios correctly, \ce{CO2} showed unrealistic abundances with respect to ice observations \citep{Boogert2015,McClure2023}. The reason for the mismatch is that, when developing the FA/\ce{CO2} reaction network according to \citet{Molpeceres2022c,Molpeceres2023,Ishibashi2024,Molpeceres2025}, i.e. slow down \ce{CO2} formation through the HOCO radical instead of \ce{CO + OH -> CO2 + H}, the abundance of \ce{CO2} decreases. The reduction is a factor 2-3 with respect to the observations. However, not including the changes totally banishes FA chemistry, which poses an even larger observational conundrum, especially in light of contemporary observations. In this Appendix we show that we can reconcile \ce{CO2} and FA abundances by simply enhancing oxygen atom diffusion in our models. In Figure \ref{fig:co2} we show the ice abundances of \ce{CO2} and FA with models using different values of $E_{\rm diff}/E_{\rm bin}$ exclusively for the O atom (in addition to the experimentally constrained value of CO, \citep{Furuya2022}), as a proxy for varying the diffusion energy. We observe that a lowering of the factor (originally 0.40; black dotted line) to the range of 0.25--0.30, in accordance with the values for CO and other closed shell adsorbates \citep{Furuya2022} allows us to reproduce observational evidence for \ce{CO2} with minimal change with respect to FA abundance for which the observational agreement is good in all cases.

The reason for changing the diffusion energy of the O atom is that, when O atom becomes more mobile on the surface the following reactions activate:

\begin{align}
   \ce{HCO + O &-> CO2 + H} \\
   \ce{CO + O &-> CO2}.
\end{align}
Because these reactions do not interfere with the FA formation network, they allow for an independent match between FA and \ce{CO2}. This finding sums to the increasing body of evidence indicating that interstellar icy \ce{CO2} forms in alternative ways to the ones traditionally considered \citep{jimenez-serra_modelling_2025} and strengthens the need for a dedicated quantum chemical or experimental study on the pair of reactions shown above.

Finally, it remains to consider whether arbitrarily varying $E_{\rm diff}/E_{\rm bin}$ is a good practice to explain the \ce{CO2} discrepancies. While certainly $E_{\rm diff}/E_{\rm bin}$ cannot be constrained in this way, there is experimental evidence to believe that many adsorbates can have an average value below 0.4--0.5 \citep{Furuya2022} contrary to what was anticipated earlier (including O atoms) \rev{\citep{Minissale2016, 2017SSRv..212....1C}}. This makes considering a lower $E_{\rm diff}/E_{\rm bin}$ justified in light of the astrochemical contemporary advances. Nevertheless, a more elegant explanation for the diffusion of O- atoms can be found when describing binding and diffusion with binding energy distributions instead of single values, where O-atoms in low binding sites will diffuse and recombine fast as shown in \citet{Furuya2024} where, unsurprisingly, icy \ce{CO2} is one of the most affected species by the inclusion of a multibinding approach.

\end{document}